\documentclass[
  aip,
  jcp,
  amsmath,amssymb,
  reprint,
]{revtex4-2}

\usepackage{graphicx}
\usepackage{listings}
\usepackage{xcolor}
\usepackage{booktabs}
\usepackage{bm}
\usepackage{textcomp}\usepackage{ulem}
\usepackage[colorlinks=true, linkcolor=blue, citecolor=blue, urlcolor=blue]{hyperref}

\definecolor{jlkw}{RGB}{0,90,160}
\definecolor{jlmacro}{RGB}{150,40,140}
\definecolor{jlcom}{RGB}{100,110,100}
\definecolor{jlstr}{RGB}{160,80,0}
\lstdefinelanguage{Julia}{
  morekeywords={function,end,for,in,if,else,elseif,while,return,using,import,
    begin,struct,mutable,const,let,do,module,true,false},
  sensitive=true,
  morecomment=[l]{\#},
  morestring=[b]",
}
\lstdefinestyle{jl}{
  language=Julia,
  basicstyle=\ttfamily\footnotesize,
  keywordstyle=\color{jlkw}\bfseries,
  commentstyle=\color{jlcom}\itshape,
  stringstyle=\color{jlstr},
  showstringspaces=false,
  columns=fullflexible,
  breaklines=true,
  frame=single,
  rulecolor=\color{black!30},
  framesep=4pt,
  xleftmargin=4pt,
  literate={@}{{\textcolor{jlmacro}{@}}}1,
}
\newcommand{\elemco}{\texttt{ElemCo.jl}}
\newcommand{\jlmol}{\texttt{JLmol}}
\newcommand{\code}[1]{\texttt{#1}}

\begin{document}

\title{\elemco{}: A Julia package for electron-correlation methods}

\author{Daniel Kats}
\email{d.kats@fkf.mpg.de}
\affiliation{Max Planck Institute for Solid State Research, Heisenbergstra\ss{}e 1, 70569 Stuttgart, Germany}

\author{Charlotte Rickert}
\affiliation{Max Planck Institute for Solid State Research, Heisenbergstra\ss{}e 1, 70569 Stuttgart, Germany}
\affiliation{Institut für Chemie, Humboldt-Universität zu Berlin,  Brook-Taylor-Straße 2, 12489 Berlin, Germany}

\author{Thomas Schraivogel}
\affiliation{Scientific Computing Center, Karlsruhe Institute of Technology,
Hermann-von-Helmholtz-Platz 1, 76344 Eggenstein-Leopoldshafen, Germany}

\author{Kristoffer Simula}
\affiliation{Max Planck Institute for Solid State Research, Heisenbergstra\ss{}e 1, 70569 Stuttgart, Germany}

\author{Denis Usvyat}
\affiliation{Institut für Chemie, Humboldt-Universität zu Berlin,  Brook-Taylor-Straße 2, 12489 Berlin, Germany}

\author{Fangcheng Wu}
\affiliation{Max Planck Institute for Solid State Research, Heisenbergstra\ss{}e 1, 70569 Stuttgart, Germany}

\date{\today}

\begin{abstract}
We present \elemco{}, an open-source Julia package for molecular electronic structure and properties
calculations with a particular emphasis on Coupled Cluster and Distinguishable Cluster methods.
The package provides a high-level, macro-based user interface which makes routine calculations
accessible to users with no prior Julia experience.
A newly developed visualizer, \jlmol{}, assists in the graphical preparation of \elemco{} input files
and displays results such as molecular orbitals.
\elemco{} offers restricted closed-shell and unrestricted variants of state-of-the-art
electron-correlation methods such as FCI, MP2, CCSD(T) as well as EOM-CCSD and the DC methods DCSD and DC-CCSDT.
In addition to traditional approaches, \elemco{} provides recently developed methods that are currently
unique to the package.
These include tensor-decomposed implementations of DCSD and DC-CCSDT (SVD-DCSD and SVD-DC-CCSDT)
as well as two-determinant and fixed-reference CC and DC methods.
For excited-state calculations, \elemco{} also offers EOM-DCSD, which is benchmarked in this work
against CC3 on the QUEST3 benchmark set, alongside EOM-CCSD.
Furthermore, both ground and excited states, including those of multireference character,
can be treated using CIPHI – an efficient selected Configuration Interaction (CI) approach employing
a CIPSI/Heat-Bath-CI-based algorithm.
Users can directly invoke internal functions from the input file, enabling them to test
and compose new methods without the need to modify \elemco{}’s source code.
Also, \elemco{} can be readily interfaced with external quantum chemistry codes through the \textsc{Fcidump} format,
for example as the high-level solver for transcorrelated Hamiltonians, periodic embedded fragments, etc.
We provide a detailed overview of \elemco{}’s methodological repertoire and discuss its technical
details and implementation, performance, interfaces, and usage.
\end{abstract}

\maketitle

\begin{figure}[t]
  \centering
  \includegraphics[width=0.30\columnwidth]{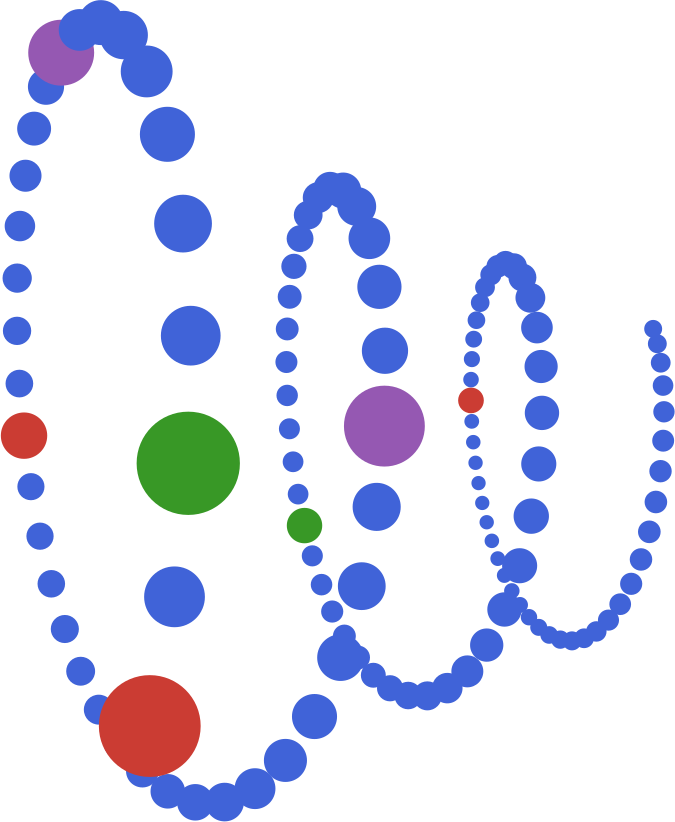}
  \caption{The \elemco{} logo.}
  \label{fig:logo}
\end{figure}

%==============================================================================
\section{Introduction}\label{sec:intro}
%==============================================================================

Quantum chemistry program packages are nowadays widely used by both theoreticians and experimentalists. 
Among wavefunction-based electron-correlation methods, Coupled Cluster (CC) theory is widely regarded as the 
gold standard of molecular electronic-structure theory because it provides systematically 
improvable and size-extensive descriptions of dynamic electron correlation.\cite{purvis1982ccsd,raghavachari1989ccsdt,bartlett2007rmp}
Thus, a large number of quantum chemistry program packages provide CC implementations, including but not limited to 
\textsc{Molpro},\cite{werner2012molpro,werner2020molpro} \textsc{Cfour},\cite{matthews2020cfour} \textsc{Orca},\cite{neese2020orca} 
\textsc{Turbomole},\cite{furche2014turbomole} 
\textsc{Mrcc},\cite{kallay2020mrcc} \textsc{NWChem},\cite{apra2020nwchem} and \textsc{Gamess}.\cite{barca2020gamess}
These packages also offer a wide range of electron-correlation methods beyond CC, including approaches designed 
to describe systems with significant static electron correlation.
These codes, typically written in Fortran or C/C++, represent decades of development and are often highly efficient. 
However, understanding their source code, adapting their functionality to specific needs, and extending them with new methods 
can be challenging, particularly for early-career scientists and experimental chemists. 
Moreover, when exploring new ideas for which only a proof of concept is initially required, the implementation process 
can be unnecessarily lengthy, even though high performance is not yet essential.

More recently, Python-based packages such as \textsc{PySCF}\cite{sun2018pyscf,sun2020pyscf} or PyBEST\cite{PyBEST2024} % and \textsc{Psi4}\cite{smith2020psi4} 
have been developed with an emphasis on code readability and ease of extension. 
In these packages, quantum-chemical building blocks exposed 
as scriptable objects have considerably lowered the barrier to method development.
Despite its productivity benefits, Python suffers from the “two-language problem”: performance-critical kernels 
are typically implemented in a compiled language, 
which creates a split between a user-friendly scripting layer and a less transparent but highly efficient core.

The Julia programming language\cite{bezanson2017julia} was designed to overcome this limitation: it combines the interactivity 
and expressiveness of a dynamic language with performance approaching that of C or Fortran, achieved 
through just-in-time compilation and an aggressive type-specializing compiler. Its multiple-dispatch paradigm makes 
it natural to write generic algorithms that operate transparently.
These features make Julia an attractive choice for quantum chemistry applications. 
For example, the \textsc{DFTK}\cite{HerbstDFTK2021} and \textsc{Fermi.jl}\cite{aroeiraFermi2022} packages 
leverage the advantages offered by Julia.

\elemco{}\cite{elemcojl} (pronounced ``elemcoil'', logo in Figure \ref{fig:logo}) was originally developed to provide a coupled-cluster (CC) solver for non-Hermitian Hamiltonians 
arising in transcorrelation \cite{boys1969calculation,schraivogel2021tc,schraivogel2023tc2}  and therefore not supported by most conventional CC codes.
The initial implementation was limited to CCSD and Distinguishable Cluster Singles and Doubles (DCSD)\cite{katsManby2013dc,kats2014dc} 
for general Hamiltonians. Over time, however, \elemco{} evolved substantially owing to Julia's combination of extensibility and high 
performance and has grown into a standalone quantum chemistry package. Consequently, a wide range of CC and DC methods for ground and excited states, 
together with mean-field, perturbative, and Configuration Interaction (CI) approaches, have been incorporated. Furthermore, numerical capabilities 
such as density fitting (DF)\cite{whitten1973coulombic, baerends1973self, dunlap1979some, vahtras1993ri, weigend2002ri}, Cholesky decomposition \cite{beebe1977simplifications,koch2003cholesky,folkestad2019efficient}, 
Laplace transform \cite{almlof1991elimination,haser1992laplace,kats2008use}, singular-value decomposition (SVD) \cite{kinoshita2003singular}, orbital localization \cite{knizia2013ibo,pipek1989localization,boys1960localization,pulay1986orbital,saebo1993local}, 
and the availability of 
Davidson \cite{shavitt1973iterative} and direct inversion in the iterative subspace (DIIS) solvers \cite{pulay1980convergence,hamilton1986direct}, among others, make \elemco{} a versatile platform 
for implementing electron-correlation methods with state-of-the-art performance.

Overall, the result is a package that is not only easy to extend but also straightforward to use. 
Calculations are initiated through intuitive macros and can be embedded in ordinary Julia scripts, 
such that a complete input file typically consists of only a handful of lines and requires no knowledge 
of Julia beyond copying a template. At the same time, because every method is implemented in pure Julia, 
users can directly invoke internal functions from the input file, enabling them to test and compose new 
methods without the need to modify \elemco{}'s source code.

It should be emphasized that, beyond providing accessible and well-established methods, \elemco{} 
includes a variety of recently developed approaches that are currently 
unique to the package, including the tensor-decomposed SVD-DCSD and SVD-DC-CCSDT methods\cite{rickert_svd_dc_ccsdt, lambie2025benchmarking}, 
the two-determinant DCSD (2D-DCSD), \cite{szalay94,schraivogel2024two} fixed-reference DCSD (FR-DCSD) and DC-CCSDT (FR-DC-CCSDT) methods, \cite{schraivogel21_dc, schraivogel2024two}
and the \mbox{CIPHI} selected CI method. Equally distinctive is the systematic support 
for non-Hermitian Hamiltonians, including $\Lambda$CCSD(T), which can be employed, for example, in 
transcorrelation. Making both these newly developed methods and established approaches freely and openly 
available constitutes a central motivation of the project.

In the remainder of this paper we describe the design and
implementation of \elemco{} (Sec.~\ref{sec:design}), survey the available methods
(Sec.~\ref{sec:methods}), the interfaces (Sec.~\ref{sec:interfaces}), 
illustrate %typical
usage (Sec.~\ref{sec:usage} and \ref{sec:phosph}), the performance (Sec.~\ref{sec:performance}), 
and close with an outlook on ongoing developments
(Sec.~\ref{sec:conclusions}).

%==============================================================================
\section{Design and implementation details}\label{sec:design}
%==============================================================================

\subsection{Installation and user interface}

\elemco{} is written entirely in Julia (version $\geq 1.9$) and can be installed with a single 
command in the Julia package manager, namely \code{Pkg.add("ElemCo")}. The only external numerical 
dependency is the \textsc{libcint} library,\cite{libcint} for integral evaluation, which is distributed 
as a precompiled binary 
%artifact 
and therefore requires no manual compilation by the user.
To ensure that calculations can be performed without any expertise in Julia, the user interface is 
built around macros of the form \code{@macroname}. An input file is an ordinary Julia 
script that loads the package via \code{using ElemCo} and then executes a sequence of macro calls. 
For example, the following input fully specifies a Hartree-Fock (HF) calculation followed 
by a CCSD(T) calculation for a nitrogen molecule with the geometry in the XYZ-format, using a 
valence double-zeta basis set:

\begin{lstlisting}[keepspaces=true]
using ElemCo
geometry = "2

  N  -0.556  0.000  0.000
  N   0.556  0.000  0.000"
basis = "vdz"
@hf
@cc ccsd(t)
\end{lstlisting}
The macros (\code{@hf}, \code{@cc}, etc.) expand into ordinary function calls at parse time,
so the convenience layer carries no run-time overhead and remains fully transparent. Because the
input is itself a Julia program, control flow, loops over geometries, parameter scans, and
post-processing can be added with the full power of the language when desired.
The \elemco{}-dedicated visualizer \jlmol{}\cite{jlmol}
-- which is essentially a wrapper around the \textsc{JSmol} library\cite{jmol,jsmol} --
 enables the graphical generation of \elemco{} input 
files and provides visualization capabilities, including the display of molecular orbitals. 
Available both as a desktop application and as a browser-based application, it offers a convenient 
alternative to manual input preparation.

Beyond the macros, every routine in \elemco{} is an ordinary, documented Julia function. Users 
seeking greater flexibility can bypass the macro interface entirely to script novel workflows 
or implement new methods by directly invoking the underlying functions, without any loss of 
capability. This ``two-tier'' design -- a declarative macro layer over a fully exposed functional 
core -- enables \elemco{} to serve both non-expert users and method developers from a single code base.

Detailed information on all macros, functions, and available options can be found on the 
\elemco{} website at \url{https://elem.co.il} and in our GitHub repository at \url{https://github.com/fkfest/ElemCo.jl}.

\subsection{Macros, options and workflow}

Table~\ref{tab:macros} provides an overview of the macros available in \elemco{} for invoking quantum chemical methods.

\begin{table*}
\caption{Overview of the macros available in \elemco{} to call quantum chemical methods.}
\label{tab:macros}
\begin{ruledtabular}
\begin{tabular}{ll}
Class & Methods \\
\colrule
Mean field        & @hf, @uhf, @dfhf, @dfuhf, @bohf, @bouhf, @dfmcscf \\
M\o{}ller-Plesset perturbation theory      & @dfmp2, @cc mp2, @dfcc mp2\\
Coupled cluster methods  & @cc [methodname], @dfcc [methodname]\\
Distinguishable cluster methods &   @cc [methodname], @dfcc [methodname]\\
Configuration interaction & @fci, @ciphi \\
Equation of motion & @cc eom-[methodname]\\
\end{tabular}
\end{ruledtabular}
\end{table*}

The macros rely on various \textit{local} variables to specify the calculation.
The system information, i.e., the molecular geometry, the basis set, or the Hamiltonian (as an \code{FCIDUMP} file\cite{knowlesDeterminant1989}),
is specified through \code{geometry}, \code{basis}, or \code{fcidump} variables.
This information, as well as various settings (including some technical settings such as the scratch directory),
is stored in an \code{ECInfo} object under the variable \code{EC} (which also lives in the local scope),
which is created automatically when the first macro is called.
The \code{ECInfo} object is passed implicitly to all subsequent macros,
so that the user does not need to manage it explicitly.
It can be reset by calling the \code{@ECinit} macro, which creates a new \code{ECInfo} object and discards the previous one.

Also the settings of the respective methods, such as convergence thresholds or the maximum number of 
iterations, are controlled through macros. Options are specified using the \code{@set} macro, either 
globally or, preferably, locally within a single calculation block, ensuring that they are automatically 
restored afterwards and cannot inadvertently affect subsequent steps, as illustrated below. 

\begin{lstlisting}
@cc ccsd begin
  #number of maximum iterations and convergence threshold
  @set cc maxit=100 thr=1.e-10 
end
\end{lstlisting}
The settings which are saved in the \code{EC::ECInfo} object are organized into thematic groups, among them \code{wf} for general wavefunction
settings, \code{scf} for self-consistent field settings, \code{cc} for CC or DC settings,
\code{eom} for EOM settings and so on.

Aside from full quantum chemical calculations and their associated settings, \elemco{} also provides 
dedicated macros for tasks such as evaluating integrals, defining dummy atoms, freezing orbitals, and 
performing orbital rotations. In addition, workflows can chain multiple macros together, with intermediate 
results such as orbitals, integrals, and amplitudes passed implicitly through a scratch directory and 
saved to or restored from wavefunction dumps using the \code{@savewf}, \code{@loadwf}, and \code{@copywf} macros.

\subsection{Tensor engine and storage}

The majority of electronic-structure methods rely on tensor contractions, and their computational 
efficiency and scalability are heavily determined by how efficiently these contractions can be performed.
In \elemco{}, these contractions are expressed through a concise Einstein-summation macro, 
called \code{@mtensor}, layered on \texttt{TensorOperations.jl},\cite{tensoroperations} 
which dispatches binary contractions to optimized \code{BLAS} routines (e.g., via \code{MKL}). 
Array views and reusable memory buffers (\texttt{Buffers.jl}) 
reduce allocations in the hot loops. 
Large intermediates that do not fit in
memory can be memory-mapped to disk transparently, so that the same code path serves both
in-core and out-of-core regimes; amplitudes and other tensors are written to and read from the
scratch directory with \code{save!}/\code{load}/\code{mmap} helpers. Cluster amplitudes follow a
virtual-first index ordering -- singles as \code{T1[a,i]} ($T^i_a$) and doubles as
\code{T2[a,b,i,j]} ($T^{ij}_{ab}$) -- so that the rate-determining contractions map directly onto
efficient \code{BLAS} calls.

A key consequence of Julia's parametric type system is that the entire correlation machinery is
written generically in the element type. The same source code therefore runs with
double-precision real (\code{Float64}) or complex (\code{ComplexF64}) arithmetic: when a complex
\textsc{Fcidump} is loaded, the element type propagates automatically through the solvers, tensor
tools, and whole quantum chemical algorithms; and the just-in-time compilation 
and multiple dispatch ensure that the support for complex-valued Hamiltonians does not incur any overhead.

\subsection{Integrals}
\label{sec:integrals}
The one- and two-electron integrals in \elemco{} are either computed through its interface to \textsc{libcint}\cite{libcint} 
or can be read from an \textsc{Fcidump} file, a wide spread format for many-body Hamiltonians.\cite{knowlesDeterminant1989} 
The latter option allows easy interfacing with other quantum chemistry software, also for 
non-symmetric Hamiltonians such as transcorrelated ones.
\elemco{} uses the density fitting (DF) approximation\cite{whitten1973coulombic, baerends1973self, dunlap1979some, vahtras1993ri, weigend2002ri} 
for \textsc{libcint}-generated integrals throughout, but the integrals are usually assembled.
From the version 0.16 (to be released soon), also standard, non-density-fitted integrals are available.
A library of standard orbital and auxiliary, density-fitting (\textsc{JKfit} and \textsc{MPfit}) 
basis sets is included in \elemco{}, obtained from the \textsc{BasisSetExchange} library\cite{pritchardNewBasis2019}.

The lifecycle of the two-electron integrals follows their origin.
\textsc{Fcidump} integrals are in the molecular-orbital (MO) basis
and are treated as user-provided data: once read, they remain the
active integral set for all subsequent correlated calculations until
they are replaced, e.g.\ by pointing the \code{fcidump} variable to a
different file. On the DF route the MO integrals are, by contrast, a
transient object: the three-index intermediates are computed via
\textsc{libcint} and the required four-index MO blocks are assembled
on demand inside the correlated calculation and deleted when it
finishes. If several calculations are to share one DF-MO integral
set, it can be generated explicitly with \code{@dfints}; an
explicitly generated set persists like a \textsc{Fcidump} one. 
Finally, the exact (non-DF) AO integrals are generated automatically by
the Hartree-Fock macros \code{@hf}/\code{@uhf}, or on explicit
request with \code{@ints} (or by setting the \code{int} option \code{df=false}),
and persist in the scratch directory for subsequent correlated calculations until the geometry or basis set is changed.
Many correlated methods then run \textit{AO-direct}: MP2, CCSD/\mbox{DCSD} and
their variants, perturbative triples, $\Lambda$ equations and
properties, EOM, and SVD-DC-CCSDT contract the AO integrals directly,
and no full set of MO integrals is ever formed. A conventional
MO-basis calculation on top of the exact AO integrals remains
available through \code{@moints} (or by setting the \code{int} option \code{ao\_direct=false}),
which transforms the integrals to
the MO basis and stores them persistently, in the same
packed representation as the \textsc{Fcidump} integrals.

The exact four-index AO integrals are stored as ``$\pm$
supermatrices''. Real AO integrals possess the full eight-fold
permutational symmetry, which would in principle permit an $n^4/8$
packing; the store is, however, deliberately built on the smaller
symmetry group that survives for complex and (to some degree) for
similarity-transformed (e.g.\ transcorrelated) integrals: the joint
exchange symmetry $v_{\mu\nu}^{\rho\sigma}=v_{\nu\mu}^{\sigma\rho}$
together with hermiticity (for non-transcorrelated integrals).
By splitting the integrals
into bra-exchange symmetric and antisymmetric combinations,
\begin{equation}
V^{\pm}_{(\mu\nu)(\rho\sigma)} \;=\;
v_{\mu\nu}^{\rho\sigma} \pm v_{\nu\mu}^{\rho\sigma},
\qquad \mu\le\nu,\quad \rho\le\sigma,
\end{equation}
the joint symmetry is transformed into two independent ones: $V^{+}$ and
$V^{-}$ are symmetric (hermitian, for complex integrals) matrices in
the packed pair space, and both index pairs $(\mu\nu)$ and $(\rho\sigma)$ become triangular.
Only the lower block triangle of each supermatrix is kept on disk (i.e., the total storage requirement is
$\approx n^4/4$), with the upper part reconstructed as the
(conjugate-)hermitian mirror at read time.
The Hartree-Fock and the AO-direct correlated methods stream this store
panel-wise through a memory map; details of the layout and of the
AO-driven contractions are given in the \elemco{} theory
documentation.

The molecular-orbital integrals are stored and addressed using physicists' notation,
$v_{pq}^{rs} = \langle pq \mid rs \rangle$. Conventional four-index integrals are kept 
in a memory-efficient packed representation: exploiting their permutational symmetry,
the last two orbital indices are combined into a single upper-triangular composite index ($r \leq s$), 
such that the $\sim n^4$ integrals are stored as a three-index array, roughly halving the storage 
requirements, utilizing the only symmetry of two-electron integrals from non-Hermitian Hamiltonians.
Whenever a particular block is requested, it is expanded back to its full four-index 
form on the fly. By contrast, density-fitted and Cholesky integrals are stored directly in their 
factorized three-index representation, $v_{p}^{qL}$, where $L$ labels the (Coulomb-metric-orthogonalized) auxiliary functions.

These sub-blocks of the integrals are retrieved through a compact, string-based interface. The functions
\code{ints1(EC,\,"ov")} and \code{ints2(EC,\,"oovv")} return the one- and two-electron integrals
over the requested orbital subspaces, where each character names a space: \code{o}/\code{v} for
occupied/virtual ($\alpha$) orbitals, \code{O}/\code{V} for their $\beta$ counterparts, \code{:}
for the full molecular-orbital space, plus further labels for doubly/singly occupied and positron
spaces. The order of the characters is significant (\code{"ov"}~$\neq$~\code{"vo"}), and the
upper/lower case of the labels selects the spin case -- so that \code{ints2(EC,\,"oovv")} returns
the closed-shell $\langle ij|ab\rangle$ block (indexed \code{[i,j,a,b]}), whereas
\code{ints2(EC,\,"oOvV")} returns the corresponding $\alpha\beta$ block. 
The \code{EC::ECInfo} object carries the orbital-space definitions.

%==============================================================================
\section{Quantum chemical methods}\label{sec:methods}
%==============================================================================

\elemco{} provides a broad range of electron-correlation methods. Table~\ref{tab:methods} 
summarizes the currently implemented approaches, 
which are described in detail in the following subsections.
The working equations of most methods can be found in the documentation of \elemco{}. \cite{elemcojl}

\begin{table*}
\caption{Overview of the quantum chemical methods available in \elemco{}. Closed-shell (spin-restricted),
restricted-open-shell (prefix R), and spin-unrestricted (prefix U) variants are available for the HF, MP2,
CC, DC, and excited-state methods as indicated. All abbreviations can be found in Section \ref{sec:methods}.}
\label{tab:methods}
\begin{ruledtabular}
\begin{tabular}{ll}
Class & Methods \\
\colrule
Mean field        & HF, UHF, DF-HF, DF-UHF, BO-HF, BO-UHF, DF-MCSCF \\
Perturbation      & DF-MP2, MP2, UMP2, SOS-LT-DF-MP2 \\
Coupled cluster   & CCSD, RCCSD, UCCSD, CCSD(T), RCCSD(T), UCCSD(T), $\Lambda$CCSD(T), $\Lambda$UCCSD(T), \\
                  & CCSDT, UCCSDT \\
Distinguishable cluster & DCSD, RDCSD, UDCSD, DC-CCSDT, UDC-CCSDT \\
Quasi-variational CC and DC & QV-CCD, OQV-CCD, QV-DCD, OQV-DCD \\
Tensor decomposed & SVD-DCSD, SVD-DC-CCSDT \\
Two-determinant & 2D-CCSD, 2D-DCSD, FR-CCSD, FR-DCSD, FR-CCSDT, FR-DC-CCSDT \\
Configuration interaction & FCI, CIPHI \\
Excited states (EOM) & EOM-CCSD, EOM-RCCSD, EOM-UCCSD, EOM-DCSD, EOM-RDCSD, EOM-UDCSD \\
Tensor networks   & DMRG (via \texttt{ITensors.jl}) \\
\end{tabular}
\end{ruledtabular}
\end{table*}

\subsection{Hartree-Fock methods and orbital localization}

\elemco{} implements standard and density-fitted Hartree-Fock in both spin-restricted 
closed-shell (HF and DF-HF) and spin-unrestricted (UHF and DF-UHF) variants. In addition to 
these canonical approaches, \elemco{} provides biorthogonal Hartree-Fock (BO-HF/BO-UHF), 
in which distinct left and right orbital sets are determined, a feature that is crucial 
for the orbital optimization in non-Hermitian Hamiltonians. For multireference problems, a density-fitted 
multiconfigurational self-consistent-field (DF-MCSCF) method employing combined first- and second-order 
orbital optimization\cite{kreplinMCSCF2020,SCI1989,augmentedHessian1981} is also available.

Once the orbitals have been obtained, localization of the occupied orbitals is supported through 
the intrinsic bond orbital (IBO),\cite{knizia2013ibo} Pipek-Mezey,\cite{pipek1989localization} 
and Foster-Boys\cite{boys1960localization} schemes, while orthogonalized projected atomic orbitals \cite{pulay1986orbital,saebo1993local} can be employed 
for the virtual space.

\subsection{M\o{}ller-Plesset perturbation theory}

Second-order M\o{}ller-Plesset perturbation theory is available in closed-shell (MP2, 
DF-MP2), and unrestricted (UMP2) variants. A scaled-opposite-spin Laplace-transform
density-fitted formulation (SOS-LT-DF-MP2) is also provided for reduced-scaling estimates.

\subsection{Coupled-cluster methods}

Coupled Cluster (CC) methods available in \elemco{} include CC with 
Singles and Doubles (CCSD),\cite{purvis1982ccsd} its perturbative triples 
extension CCSD(T),\cite{raghavachari1989ccsdt} the corresponding $\Lambda$CCSD(T) 
variant,\cite{kucharskiNoniterativeEnergy1998} and Coupled Cluster with iterative triples (CCSDT).
All aforementioned methods are implemented in closed-shell, restricted open-shell (R),\cite{knowlesCoupled1993,knowlesErratum2000a} 
and spin-unrestricted (U) variants.
For the implementation of CCSDT and UCCSDT, the working equations were automatically 
generated using our second-quantization program \textsc{Quantwo},\cite{quantwo} 
which derives the working equations symbolically, factorizes the resulting contractions, 
and emits the corresponding Julia tensor-contraction kernels.

The CCSD factorization is partially based on Ref.~\citenum{hampelComparison1992}, i.e., the most
expensive step of the CCSD iteration is performed in the full (MO or AO) basis.
In the AO-direct implementation, the most expensive contraction is performed in the AO basis,
directly contracting the $\pm$ combinations of AO integrals with the $\pm$ combinations of $T_2$ amplitudes,
which reduces the cost of the contraction by a factor of 2. The remaining contractions are performed in the MO basis.

\subsection{Distinguishable Cluster methods}
\label{subsec:distinguishable}
A central feature of \elemco{} is the availability of Distinguishable Cluster (DC) 
approximations\cite{katsManby2013dc,kats2014dc} to all implemented CC methods and 
beyond. The DC approximation is obtained from CCSD by removing quadratic exchange 
terms from the doubles-amplitude equations and rescaling a few other terms. 
This modification substantially improves the description of closed- and open-shell 
reactions, atomization energies, ionization potentials, electron affinities, 
interaction energies, equilibrium geometries, spectroscopic constants, systems 
with multireference character, and bond-breaking processes, while simultaneously 
restoring exactness for two-electron systems and preserving size extensivity, 
orbital-rotation invariance, and particle-hole symmetry.
\cite{kats2015dcf12,kats2018dc,katsImprovingDistinguishable2018,katsParticle2018}

\elemco{} provides DCSD together with its restricted open-shell and unrestricted variants 
(RDCSD and UDCSD), as well as iterative triples extensions\cite{kats19_dc,rishi19,schraivogel21_dc} 
in the form of DC-CCSDT and UDC-CCSDT. %\cite{kats2015dcf12,kats2018dc} 
As for the CCSDT and 
UCCSDT implementations, the corresponding working equations for their DC
counterparts were generated automatically using our second-quantization program \textsc{Quantwo}.\cite{quantwo} 

\subsection{Quasi-Variational Coupled Cluster and Distinguishable Cluster methods}
Quasi-variational coupled-cluster theory \cite{QVCCD,blackStatistical2018}
is a Hermitian variant of CC theory that incorporates some degree of variationality into
the otherwise non-variational CC ansatz.
Both quasi-variational coupled-cluster doubles (QVCCD)\cite{QVCCD} and quasi-variational distinguishable 
cluster doubles (QVDCD)\cite{QVDCD} are implemented in \elemco{} in their closed-shell formulations, 
together with their orbital-optimized variants, OQVCCD and OQVDCD.

\subsection{Tensor-decomposed methods}
\label{subsec:tensor_decomposed}

To extend high-accuracy methods to larger systems, \elemco{} implements 
tensor-decomposed DC methods based on a singular-value decomposition (SVD) 
of the cluster amplitudes. For SVD-DC-CCSDT,\cite{rickert_svd_dc_ccsdt} the 
triples amplitudes are represented in a compact SVD basis,
\begin{equation}
T^{ijk}_{abc} \approx T_{XYZ} U^{iX}_a U^{jY}_b U^{kZ}_c,
\end{equation}
where $T_{XYZ}$ is a small, fully symmetric three-index core tensor and 
the compression coefficients $U^{iX}_a$ connect the SVD indices $X$, $Y$, and $Z$ to the 
occupied ($i,j,k,\ldots$) and virtual ($a,b,c,\ldots$) orbital spaces. 
The triples amplitudes are optimized directly in this SVD-basis.
Combined with DF, this reduces the nominal computational 
scaling of DC-CCSDT from $O(N^8)$ to $O(N^6)$, as the number of SVD 
functions is expected to grow linearly with system size. The SVD basis is constructed 
from an auxiliary CC3-type triples density, using a threshold to retain only 
the most important singular vectors based on the magnitude of the
corresponding singular values. 
An optional hybrid 
correction, $E_{\mathrm{SVD-DC\text{-}CCSDT\text{+}}} \approx E_{\mathrm{SVD\text{-}DC\text{-}CCSDT}} +
(E_{\mathrm{CCSD(T)}} - E_{\mathrm{SVD\text{-}CCSD(T)}})$, 
accelerates convergence with respect to the SVD threshold.
An analogous SVD-DCSD method is also implemented, in which the doubles amplitudes are 
represented in an SVD basis.
Currently, SVD-DCSD is available as a pilot implementation and 
is scheduled for inclusion in the next \elemco{} release.

\subsection{Two-determinant and fixed-reference methods}

For challenging spin states in which single-determinant wavefunction ansätze become inadequate, 
\elemco{} offers two-determinant variants of CCSD and DCSD (2D-CCSD and 2D-DCSD),\cite{szalay94,schraivogel2024two} 
together with fixed-reference methods (FR-CCSD and FR-DCSD) and their triples extensions FR-CCSDT 
and FR-DC-CCSDT.\cite{schraivogel21_dc,schraivogel2024two} The defining feature of the fixed-reference approaches is that a selected reference 
$\alpha\beta$ doubles amplitude remains fixed during the iterative optimization.
These methods can be applied to biradical systems to enforce the correct spin state.

\subsection{Configuration interaction: FCI and CIPHI}\label{sec:ciphi}
In addition to an efficient full configuration interaction (FCI) implementation, \elemco{} also offers its own selected CI method, termed CIPHI (Configuration Interaction using a Perturbative/Heat-bath Iterative selection), which becomes an effective alternative to FCI for active spaces that are too large for the latter.
%is described in the 
%next subsection.
%\subsubsection{CIPHI: a selected configuration-interaction method}\label{sec:ciphi}
%For active spaces that are too large for FCI, \elemco{} provides its own selected 
%CI method, termed CIPHI, which
%It combines perturbative and heat-bath-based iterative selection strategies.
As its name suggests, CIPHI incorporates 
ideas from both CIPSI\cite{huron73_cipsi} and heat-bath CI (HCI).\cite{holmes16_hci,sharma_semistochastic_2017} 
Unlike the hybrid CIPSI/HCI approach of Ref. \onlinecite{Ugandi2023},
we employ a determinantal basis rather than configuration state functions,
which allows for the perturbative selection criterion to be evaluated with negligible computational cost.
CIPHI is similar in spirit to ASCI from Ref. \onlinecite{Tubman2020},
and differs primarily in the selection and prescreening algorithm, which we summarize below.
%Since, to the best of our knowledge, the method has not been described elsewhere, we 
%ummarize it briefly in the following.

CIPHI iteratively selects the most important determinants for the CI expansion. 
Each macro-iteration consists of four steps:
(i)~the pre-selection of new determinants based on the heat-bath criterion,
(ii)~the final perturbative selection of new determinants, 
(iii)~construction (extension) of the CI Hamiltonian in the selected determinant space, and
(iv)~diagonalization of the Hamiltonian in the enlarged determinant space.
The procedure is repeated until either 
a target expansion size is reached or the CI energy has converged.
At the end, a perturbative correction is evaluated.

\paragraph*{(i) Heat-bath preselection.}
For every selected determinant with CI coefficient $c_I$, singly and
doubly excited candidate determinants $|J\rangle$ satisfying the
heat-bath criterion
\begin{equation}
  |H_{JI}\, c_I| > \epsilon_h
\end{equation}
are generated, where $\epsilon_h$ is a screening threshold (by default $\epsilon/10$, with
$\epsilon$ the master selection threshold). To make this step
efficient, the double-excitation matrix elements are precomputed once and stored, for each ordered
orbital pair, in lists sorted by decreasing $|H|$ (a precomputed-heat-bath, PCHB, data structure
following Holmes et al.\cite{holmes16_hci}). The sorted lists allow
the generation loop to stop as soon as $|H_{JI}|$ falls below the adaptive threshold
$\epsilon_h/|c_I|$, so that the cost of proposing candidates scales with the number
of important connections rather than with the full size of the excitation manifold.
An additional instantaneous coefficient threshold $\epsilon_c$ can be used to discard candidates
with negligible instantaneous weight,
\begin{equation}
  \bigl| c_J^{\text{inst}I} \bigr| = \bigl| \frac{H_{JI}c_I}{E_{\mathrm{var}} - H_{JJ}} \bigr| < \epsilon_c,
  \label{eq:inst}
\end{equation}
on the fly (this is however optional). The denominator in Equation \eqref{eq:inst} is efficiently obtained using the relation
\begin{equation}
  E_{\mathrm{var}} - H_{JJ} = E_{\mathrm{var}} - H_{II} - \Delta H_{JJ,II},
\end{equation}
where $\Delta H_{JJ,II} = H_{JJ} - H_{II}$ is the difference between the diagonal
Hamiltonian matrix elements of the selected and candidate determinants,
which only depends on the excitation information (and the diagnonal of the Fock matrix of $I$),
and therefore can be precomputed and stored along with the PCHB data structure.

The candidate determinants that survive the heat-bath preselection are stored as a dictionary,
with the candidate determinants as keys and the sums of the $c_J^{\text{inst}I}$ and $H_{JI}c_I$ (for PT2 correction only)
contributions from all connected selected determinants as values. This allows for efficient
perturbative selection in the next step, as well as for an estimate of the perturbative correction
at the end of the selection process.
\paragraph*{(ii) Perturbative refinement.}
The preselected determinants are then ranked by their first-order (Epstein-Nesbet) perturbative
coefficient,
\begin{equation}
  \bigl|c_J^{(1)}\bigr|^2 \approx
  \frac{\bigl| \sum_{I\,:\,|H_{JI}c_I| > \epsilon_h} H_{JI}\, c_I \bigr|^2}
       {\bigl(E_{\mathrm{var}} - H_{JJ}\bigr)^2}=\bigl|\sum_{I}c_J^{\text{inst}I}\bigr|^2,
\end{equation}
and only those with $\bigl|c_J^{(1)}\bigr|^2 \ge \epsilon_p^2$ are chosen,
where $\epsilon_p$ is by default equal to $\epsilon$.
The sum runs only over the variational determinants that survived the heat-bath
criterion and is obtained virtually for free from the preselection step,
as the sum of instantaneous contributions $c_J^{\text{inst}I}$.
The combination of both selection steps yields a compact 
determinant space in which CI can be performed while retaining the 
determinants that contribute most significantly to the energy.

\paragraph*{(iii) CI Hamiltonian construction.}
The CI Hamiltonian in the selected determinant space is constructed by evaluating the matrix
elements between the selected determinants.
The connectivity of the selected determinants is determined similarly to Ref.~\citenum{sharma_semistochastic_2017},
by introducing helper dictionaries of same-$\alpha$-string and same-$\beta$-string determinants,
which allow for efficient identification of determinants connected by same-spin excitations,
and additionally a single-$\alpha$-excitation dictionary, which stores for all existing $\alpha$-strings
all $\alpha$-strings connected to them by a single excitation, allowing for efficient
identification of determinants connected by opposite-spin excitations.

\paragraph*{(iv) Diagonalization.}
The CI Hamiltonian in the selected determinant space is diagonalized
directly for small spaces ($\lesssim 10^3$ determinants) or with a Davidson solver for larger ones.
An optional small-space pre-diagonalization
provides a good initial guess, and several roots can be targeted simultaneously
(\code{nstates}~$>1$) for excited states; in the multi-state case the selection uses a
state-maximum criterion, i.e. a determinant is added if its perturbative coefficient is
significant for any of the targeted states, which avoids biasing the space toward the
ground state. 

The entire cycle, consisting of steps (i), (ii), (iii) and (iv), is repeated until either the determinant 
count reaches the specified target or the energy has converged to the requested tolerance.
\paragraph*{Perturbative correction and variants.}
Once the variational space has converged, a second-order Epstein-Nesbet correction,
\begin{equation}
\Delta E_{\mathrm{PT2}} = \sum_{K \notin \mathcal{V}}
\frac{\bigl(\sum_{I \in \mathcal{V}} H_{KI}, c_I\bigr)^2}{E_{\mathrm{var}} - H_{KK}},
\end{equation}
is evaluated, where $K$ runs over determinants outside the selected determinantal space. This is
computed by the same heat-bath machinery as the selection, but with a tighter threshold
$\epsilon_{\mathrm{pt2}}$ (and an adaptive variant $\epsilon_{\mathrm{pt2}}/|c_I|$); a further
instantaneous screening discards negligible contributions before they are fully assembled, which
bounds the memory cost and also yields an uncertainty estimate of the correction. An uncontracted
MP2 form and a renormalized variant, $E_{\mathrm{PT2}} \rightarrow E_{\mathrm{PT2}}/(1+\lVert
T_2\rVert^2)$, are also available. CIPHI calculations can be stored to and restarted from a portable
\textsc{TREXIO} dump -- useful for tightening thresholds, computing the PT2 correction on a
pre-converged space (a PT2-only mode), or storing the states of a multi-state calculation
separately.

Like the rest of \elemco{}, the FCI and CIPHI solvers are formulated generically in the element
type and do not assume Hermiticity, so both also operate for complex-valued and non-Hermitian Hamiltonians.

\subsection{Molecular properties}\label{sec:properties}

Beyond energies, \elemco{} can evaluate one-electron molecular properties at the CC, DC, 
and CI levels. From the one-particle reduced density matrix (1-RDM), obtained by solving 
the associated CCSD/DCSD Lagrange-multiplier equations, the package can compute properties 
such as dipole moments. Molecular properties can be evaluated with (U)CCSD and (U)DCSD by setting 
\code{@set cc properties=true}. The corresponding properties are likewise available for FCI and 
CIPHI wavefunctions.

Furthermore, natural orbitals and their occupation numbers can be obtained by diagonalizing 
the 1-RDM. These natural orbitals and occupations can be written to a \textsc{TREXIO} dump 
for subsequent reuse, for example in active-space selection procedures.

\subsection{Excited states}
\label{subsec:excited_states}
In \elemco{}, excited states, in addition to FCI or CIPHI (see above), can be computed using equation-of-motion CC
(EOM-CC) theory,\cite{stanton1993eomcc,koch1990lr} implemented at the  EOM-CCSD level 
and its DC counterpart EOM-DCSD. Both methods are available in 
closed-shell, restricted open-shell (R), and spin-unrestricted (U) variants 
(EOM-RCCSD/EOM-UCCSD and EOM-RDCSD/EOM-UDCSD). The corresponding trial vectors and 
Hamiltonian matrices are formulated for general numeric types, making excited-state 
calculations for non-Hermitian and complex-valued Hamiltonians equally accessible. 
%In addition, FCI and CIPHI can also be used to obtain excited states.

\begin{figure*}[t]
  \centering
  \includegraphics[width=\textwidth]{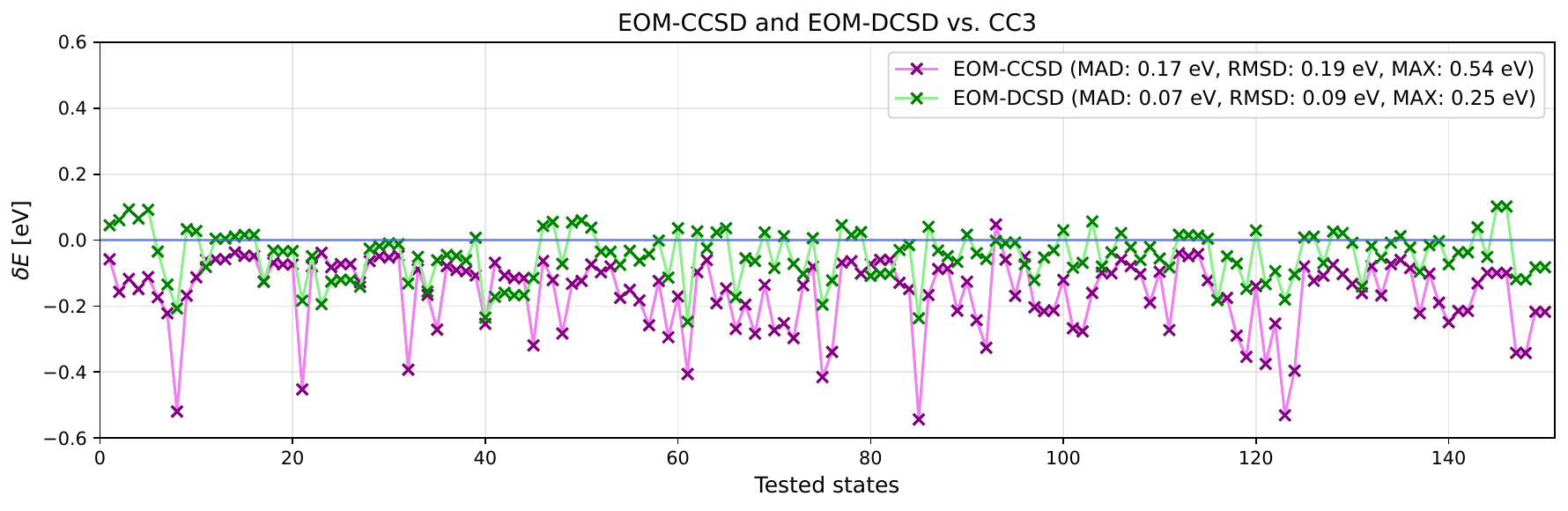}
  \caption{Benchmark of 150 vertical single-excitation energies computed 
  with EOM-DCSD and EOM-CCSD relative to CC3 for the QUEST3 benchmark 
  set,\cite{loos2020mountaineering,veril2021questdb} which comprises 
  medium-sized molecules containing four to six non-hydrogen atoms. 
  All calculations employ the augmented triple-zeta basis. 
  MAD, RMSD, and MAX denote the mean absolute deviation, 
  root-mean-square deviation, and maximum deviation, respectively.}
  \label{fig:EOM_vs_CC3}
\end{figure*}

To assess the accuracy of the EOM-DCSD implementation, we present a benchmark 
against CC3 for the QUEST3 benchmark set \cite{loos2020mountaineering,veril2021questdb} of vertical excitation 
energies. The results are shown 
in Figure~\ref{fig:EOM_vs_CC3}. 
All EOM-DCSD results obtained in this work, together with the EOM-CCSD 
and CC3 benchmark data from QUEST3, are provided in the Supplementary 
Information, including the ordering of molecules and excited states used 
in the figure.
The QUEST3 set comprises medium-sized molecules containing four to six 
non-hydrogen atoms. We consider 150 states with dominant single-excitation character for which 
both EOM-CCSD and CC3 reference data are available in the augmented 
triple-zeta basis.\cite{dunning1989gaussian,weigend2002efficient}
Consistent with previous studies,\cite{rishi2016assessing,rishi2017excited,
rishi2023dark} EOM-DCSD outperforms EOM-CCSD for the vast majority of states. 
Of the 150 excitations considered, EOM-DCSD performs better for 133 states, 
both methods perform equally well for three states, while EOM-DCSD yields 
larger deviations from CC3 than EOM-CCSD for only 14 states.
The mean absolute deviation (MAD) and root-mean-square deviation (RMSD), 
also shown in Figure~\ref{fig:EOM_vs_CC3}, demonstrate that EOM-DCSD reduces the error (with respect to the CC3 reference)
by more than a factor of two relative to EOM-CCSD.
For EOM-CCSD, the largest deviation is observed for the 
$^{1}B_g(\mathrm{V}; n \rightarrow \pi^\ast)$ state of glyoxal, 
amounting to 0.54 eV. For the same state, the EOM-DCSD deviation is 
only 0.24 eV, i.e., less than half that of EOM-CCSD.
For EOM-DCSD, the largest deviation is observed for the 
$^{1}B_{1g}(\mathrm{V}; n \rightarrow \pi^\ast)$ state of pyrazine, 
with an error of 0.25 eV. For this state, the corresponding EOM-CCSD 
deviation is 0.41 eV, which is almost twice as large. 
It should also be noted that, in addition to being more accurate, 
EOM-DCSD is slightly less computationally expensive than EOM-CCSD. 
Like EOM-CCSD, EOM-DCSD formally scales as $\mathcal{O}(N^6)$; however, 
because certain terms in the residual equations are omitted, the 
computational prefactor is smaller.

\subsection{Non-Hermitian and Complex-Valued Hamiltonians}\label{sec:nonherm}

The majority of methods in \elemco{} do not assume Hermiticity of the Hamiltonian and 
can therefore be applied directly to non-Hermitian and to complex-valued Hamiltonians.
This capability enables electronic-structure calculations 
based on integrals supplied through a generalized \textsc{Fcidump} file, making \elemco{} suitable 
for non-Hermitian theories such as transcorrelated methods.\cite{boys1969calculation,schraivogel2021tc,schraivogel2023tc2,christlmaierXTC2023}
In particular, the xTC Hamiltonian (transcorrelated Hamiltonian excluding explicit three-body terms)
can be generated with the \textsc{(py)TCHint} program,\cite{tchint}, and the resulting integrals
can be read into \elemco{} for subsequent correlation calculations.
The non-hermiticity of the Hamiltonian is indicated by setting \code{ST=1} (i.e., Similarity-Transformed)
in the header of the \textsc{Fcidump} file.

\subsection{Mixed electron-positron systems}
\label{sec:positrons}

\elemco{} can treat mixed electron--positron systems at the density-fitted
Hartree--Fock (DF-HF) and second-order M{\o}ller--Plesset (MP2) levels within
a restricted formalism. A positron is added to the system with
\code{@set wf npositron=1}. Only a single positron is currently supported, so
no positron--positron integrals arise.

For interfacing with external solvers, \elemco{} can write and read \textsc{MOLDEN} and
\textsc{FCIDUMP} files for these systems. The \textsc{MOLDEN} files include an additional field
indicating the presence of a positron and list the positron orbitals separately
from the electronic ones; in the \textsc{FCIDUMP} files, a line of zeroes separates the
electron--positron integrals from the purely electronic ones. These extensions
make it possible to couple positron solvers with other codes, such as the
variational and diffusion quantum Monte Carlo (VMC and DMC) methods of
\textsc{casino}.

%==============================================================================
\section{Interfaces and interoperability}\label{sec:interfaces}
%==============================================================================

\begin{figure*}[t]
  \centering
  \includegraphics[width=\textwidth]{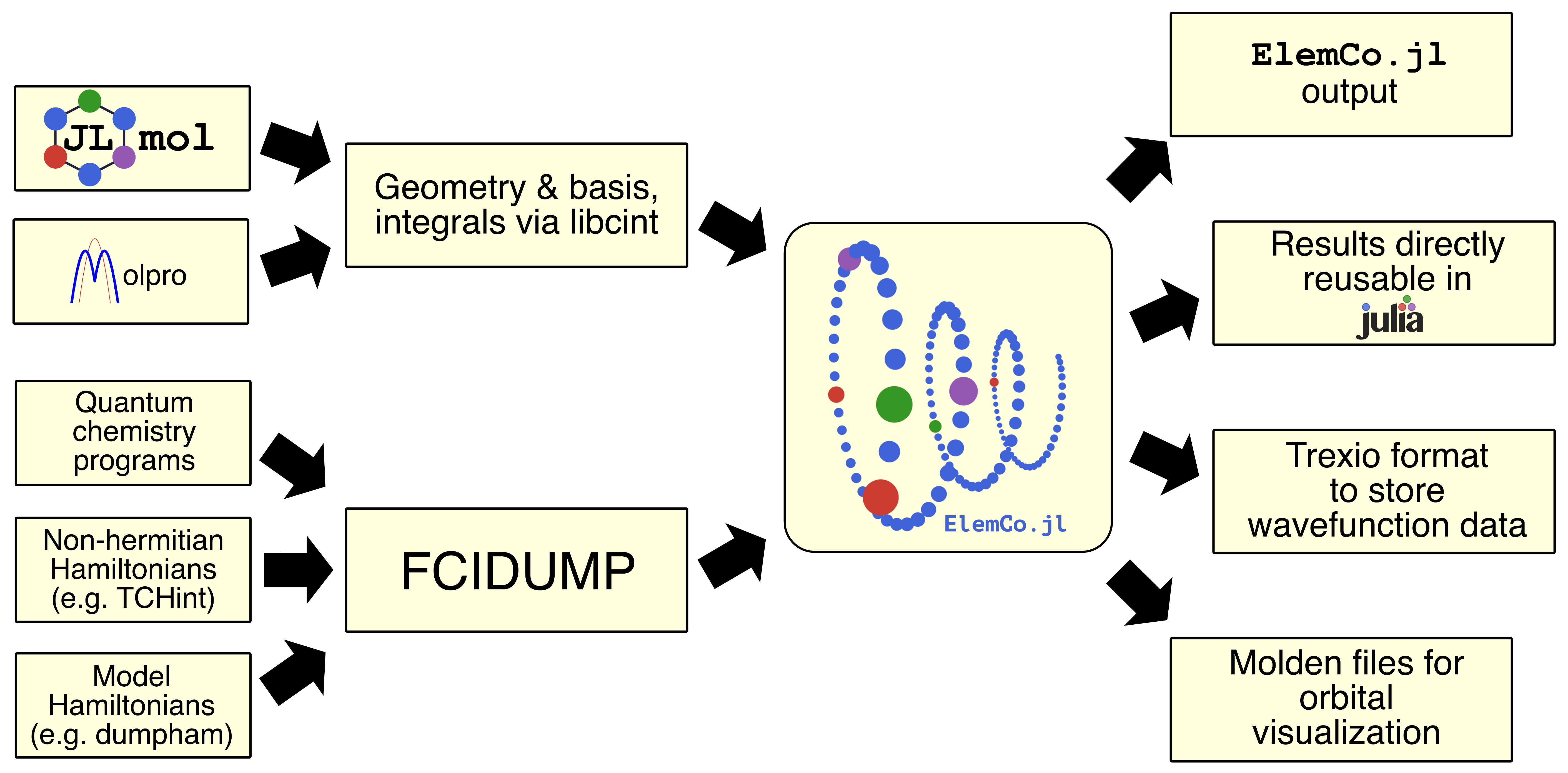}
  \caption{Schematic overview of the input and output options available in 
  \elemco{}, including supported interfaces and interoperability with external 
  software.}
  \label{fig:input_output_scheme}
\end{figure*}

\elemco{} is designed to interoperate with external quantum chemistry programs. 
A schematic overview of how \elemco{} can be used together with other programs 
and how data can be exchanged between them is shown in 
Figure~\ref{fig:input_output_scheme}.

For generating inputs in which the molecular geometry and basis set are specified 
and the corresponding integrals are evaluated via \textsc{libcint}, \jlmol{} and 
\textsc{Molpro} provide interfaces, although their use is entirely optional. 
Integrals and Hamiltonians can be exchanged 
through the \textsc{Fcidump} format, which is routinely supported by many quantum 
chemistry programs. The \textsc{Fcidump} interface also supports the transfer of 
non-Hermitian Hamiltonians, including those used in transcorrelation and generated, 
for example, with \textsc{TCHint},\cite{tchint} as well as model Hamiltonians generated with \textsc{DumpHam}.\cite{dumpham}
The extended \textsc{Fcidump} format can also store the integral values in a binary \textsc{NPY} format 
using the \elemco{}'s storage convention, section \ref{sec:integrals},
which is much more efficient for large systems than the standard ASCII format and can 
directly be read into \elemco{}.

Regarding the output, in addition to the standard text output summarizing the 
most important details and results of the quantum chemical calculations, all 
results are directly reusable in Julia. In particular, computed quantities 
can be stored as Julia objects and can therefore be processed further within user 
scripts. Wavefunction data such as orbitals and amplitudes can be written to 
portable dump files in the \textsc{TREXIO} format,\cite{posenitskiy2023trexio} 
enabling interchange with other codes. Moreover, \elemco{} can generate 
\textsc{Molden} files for orbital visualization, which can be viewed either 
directly in \jlmol{} or in other visualization programs.

Finally, \elemco{} exposes an interface to the 
\texttt{ITensors.jl}/\texttt{ITensorMPS.jl} libraries\cite{fishman2022itensor} 
for density-matrix renormalization-group (DMRG)\cite{white1992dmrg,chan2011dmrg} 
calculations.

%==============================================================================
%\section{Illustrative usage}\label{sec:usage}
%==============================================================================

\section{Representative examples}\label{sec:usage}
In this section, we present a few complete \elemco{} input examples.
Note that the examples are not meant to be exhaustive, but rather to illustrate the
basic usage of \elemco{} and its macro interface.
Besides, the program is under active development, and the input syntax or default settings
may change in future releases.  For a complete up-to-date list of available
options and their default values, please refer to the \elemco{} documentation.\cite{elemcojl}

\paragraph*{DCSD from a geometry.}
In this example, a molecular geometry and a basis set are specified. 
\elemco{} first performs a density-fitted Hartree-Fock calculation and 
subsequently a DCSD calculation.
Unless explicitly specified otherwise (here, the unit \textsc{bohr} is given 
explicitly), Bohr is used as the default unit for molecular geometries. Alternatively, 
geometries can be specified in Angstrom by writing \textsc{angstrom} instead of 
\textsc{bohr}, or by using the XYZ format (first line: number of atoms; second 
line: comment or empty line; subsequent lines: atom type and Cartesian coordinates in Angstrom).
The basis set can be specified as a dictionary, as shown below, containing entries 
for the atomic orbital basis (\textsc{ao}), the HF auxiliary basis (\textsc{jkfit}), and 
the correlation auxiliary basis (\textsc{mpfit}). Alternatively, only the AO 
basis can be provided, for example through \texttt{basis = "name\_of\_basis"}, 
in which case suitable \textsc{jkfit} and \textsc{mpfit} basis sets are selected 
automatically. For instance, specifying \texttt{basis = "vdz"} selects auxiliary 
basis sets at the valence-double-zeta level.
The use of \texttt{@print\_input} is optional.

\begin{lstlisting}[keepspaces=true]
using ElemCo
@print_input
geometry = "bohr
    O   0.000000   0.000000  -0.123995
    H1  0.000000   1.431132   0.984106
    H2  0.000000  -1.431132   0.984106"
basis = Dict("ao"    => "cc-pVDZ",
             "jkfit" => "cc-pvtz-jkfit",
             "mpfit" => "cc-pvdz-mpfit")
@dfhf
@cc dcsd
\end{lstlisting}

\paragraph*{CCSD(T) from an \textsc{Fcidump} file.}
In the following example, an \textsc{Fcidump} file named \textsc{H2O.FCIDUMP} 
is read and used as the input Hamiltonian for a CCSD(T) calculation. The energies 
from the CCSD(T) calculation are stored in the ordered dictionary \texttt{ccsd\_t\_energies}
and can be directly accessed in further Julia code.
\begin{lstlisting}
using ElemCo
fcidump = "H2O.FCIDUMP"
ccsd_t_energies = @cc ccsd(t)
\end{lstlisting}

\paragraph*{EOM-DCSD with a local option block.}
In this example, a molecular geometry is read from the file \texttt{"geometry.xyz"}. 
Subsequently, a HF calculation is performed, followed by an EOM-DCSD calculation. 
Options can be confined to a single method call so that they do not affect subsequent 
calculations. Here, the local option block is used to specify the number of EOM-DCSD 
states to be computed:
\begin{lstlisting}
using ElemCo
@print_input
geometry = "geometry.xyz"
basis = "avtz"
@hf
@cc eom-dcsd begin
  @set eom nstates=5
end
\end{lstlisting}

\paragraph*{SVD-DC-CCSDT calculation.}
For the molecular geometry specified in \texttt{"geometry.xyz"} and a 
valence-double-zeta basis set, \elemco{} first performs a DF-HF calculation 
followed by an SVD-DC-CCSDT calculation. To assess convergence with respect 
to the size of the SVD basis, as controlled by the decomposition threshold 
$\varepsilon_{T3}$, the option \texttt{@set cc ampsvdtol} can be used to 
tighten the truncation threshold. The default value is currently \texttt{1.e-5}.
\begin{lstlisting}
using ElemCo
geometry = "geometry.xyz"
basis = "vdz"
@dfhf
@cc svd-dc-ccsdt begin
  @set cc ampsvdtol = 1.e-6
end
\end{lstlisting}

\paragraph*{Scripting and invoking internal functions.}
In addition to the macro interface, \elemco{} allows users to call internal functions 
directly from the input file. In the example below, the doubles amplitudes \texttt{T2} 
from a DCSD calculation are loaded together with the occupied-occupied-virtual-virtual 
integral block \texttt{oovv}. These quantities are then contracted to form the closed-shell
doubles contribution to the correlation energy using the \texttt{@mtensor} macro, which
follows Einstein summation conventions:
\begin{lstlisting}
using ElemCo
using ElemCo.TensorTools     # ints1, ints2, @mtensor
fcidump = "FCIDUMP"
@cc dcsd
T2 = @loadfile cc_amplitudes_2
oovv = ints2(EC, "oovv")     # <ij|ab>, indexed [i,j,a,b]
@mtensor E = (2*T2[a,b,i,j] - T2[b,a,i,j])*oovv[i,j,a,b]
\end{lstlisting}
Any other quantity can be assembled in the same way: the required integral blocks can 
be obtained via \code{ints1} and \code{ints2}, amplitudes can be loaded from disk, and 
the resulting expressions can be evaluated with \code{@mtensor}. This flexibility makes 
prototyping and testing new methods on top of \elemco{} straightforward.

%==============================================================================
\section{Performance}\label{sec:performance}
%==============================================================================

In the following, the computational performance of \elemco{} is assessed using 
representative calculations on the phenol molecule consisting of 13 atoms and 50 electrons. The molecular geometry of 
phenol is provided in the Supplementary Material.
The calculations employ an augmented triple-zeta basis (aug-cc-pVTZ) \cite{dunning1989gaussian, weigend2002efficient} comprising 460 
atomic orbitals, 1003 auxiliary functions in the \textsc{JKfit} basis and 1017 auxiliary functions in the \textsc{MPfit} basis.
The DF-HF energy is -305.670426 hartree, and the corresponding 
running time is less than 10 seconds, making its cost negligible compared to that of the subsequent correlated calculations (apart from DF-MP2, which is even faster than DF-HF).

With the frozen core approximation, the correlation 
calculations involve 18 occupied orbitals and 435 virtual orbitals.
Table~\ref{tab:performance} summarizes the timings and energies obtained with 
different electron-correlation methods. 
As expected, MP2 is by far the fastest method. DCSD and CCSD are comparable in the overall cost, with DCSD being slightly faster as it omits some computationally expensive exchange diagrams, as discussed in Section~\ref{subsec:distinguishable}. As expected, the DCSD energy is closer to the 
higher-level CCSD(T) and SVD-DC-CCSDT values, as has been consistently observed for this method.\cite{katsManby2013dc,kats2014dc,kats2015dcf12}
For the present system, the perturbative (T) is approximately four times more expensive than DCSD or CCSD.

Despite containing iterative triples amplitudes, with $\varepsilon_{T3}=10^{-5}$ the wall time of the SVD-DC-CCSDT calculation is of the same order of magnitude as the non-iterative (T) correction. 
This efficiency is a consequence of the tensor-decomposed approach described 
in Section~\ref{subsec:tensor_decomposed}, which substantially reduces the computational scaling and the prefactor compared to conventional DC-CCSDT.

\begin{table}
\caption{Timings and correlation energies obtained with different electron correlation 
methods for phenol (aug-cc-pVTZ, 36 correlated electrons in 453 orbitals). Here, $t$ denotes the total wall time required 
by the respective correlation method, while $t_{\mathrm{iter}}$ is the 
average time per iteration. The time given for CCSD(T) corresponds only to the evaluation of the (T) correction.
The number of iterations is given by 
$N_{\mathrm{iter}}$, and $E_{\mathrm{cor}}$ denotes the correlation 
energy. All calculations were performed on 12-core Intel Xeon Gold
6128 nodes with a rate of 3.40 GHz.}
\label{tab:performance}
\begin{ruledtabular}
\begin{tabular}{lllll}
Method & $t$, s & $N_{\rm iter}$ & $t_{\rm iter}$, s & $E_{\rm cor}$, E$_h$ \\
\colrule
DF-MP2   & 2 & - & - & -1.205019 \\
DCSD  & 1456 & 13 & 112 & -1.259989 \\
CCSD  & 1526 & 12 & 127 & -1.223712 \\
CCSD(T)  & 5768 & - & - & -1.287759 \\
SVD-DC-CCSDT & 6344 & 12 & 346 & -1.284869 \\
\end{tabular}
\end{ruledtabular}
\end{table}

\section{Illustrative application}\label{sec:phosph}
%  to a Phosphorene Vacancy via the FCIDUMP Interface}\label{sec:phosph}

In this prototypical application, we demonstrate how the FCIDUMP interface to \elemco{} can be used to study complex systems at a high level of quantum-chemical theory. Specifically, we address the formation energy of a negatively charged vacancy in phosphorene, with particular focus on the post-CCSD(T) contribution, as a follow-up to the CCSD(T) benchmark of Ref. \onlinecite{Phosph_2026}.

Phosphorene is a single two-dimensional layer that stacks to form bulk black phosphorus. A highly accurate quantum-chemical description of black phosphorus is of particular interest in the context of a recent controversy concerning discrepancies between CCSD(T) and quantum Monte Carlo (QMC) calculations of intermolecular interactions in large dimers.\cite{al-hamdani:2021,lambie:2024,lambie:2026,Schaefer:25} Interestingly, a possibly related effect has also been observed for the exfoliation energy of black phosphorus: despite overall good agreement between the two approaches, CCSD(T) predicts\cite{schutz2017} a slightly stronger binding than QMC.\cite{ShulenburgerTomanek_QMC_BP} Here, we focus not on exfoliation but rather on the formation energy of a negatively charged vacancy. Nevertheless, the results of Ref. \onlinecite{Phosph_2026} demonstrate that van der Waals dispersion makes a sizable contribution to the vacancy formation energy, suggesting that an estimate of the post-CCSD(T) correction provides important additional information.

We calculate the defect formation energy using the recently developed aperiodic defect model (ADM)\cite{ADM_2025} as implemented in the \textsc{Cryscor} code.\cite{usvyat18} In this model, a fragment containing the defect is embedded in the periodic Hartree-Fock mean field of the pristine crystal. The fragment-environment partitioning is based on the localized orbitals of the pristine structure and does not require cutting out a physical fragment. The ADM offers several key advantages over conventional supercell calculations. In particular, it does not periodically repeat the defect image, as in the supercell approach, and thus avoids unphysical interactions between defect images. This is especially important for charged defects such as the one considered here. Another important feature of the ADM is that the embedding field is included in the fragment's one-electron Hamiltonian, effectively reducing the problem to a molecule-like one. The matrix elements of the resulting effective fragment Hamiltonian in the basis of the fragment HF orbitals can then be written in FCIDUMP format and passed to \elemco{}. A detailed description of the ADM and its formalism can be found in Ref. \onlinecite{ADM_2025}.

\begin{figure}
\centering
\includegraphics[width=0.45\textwidth]{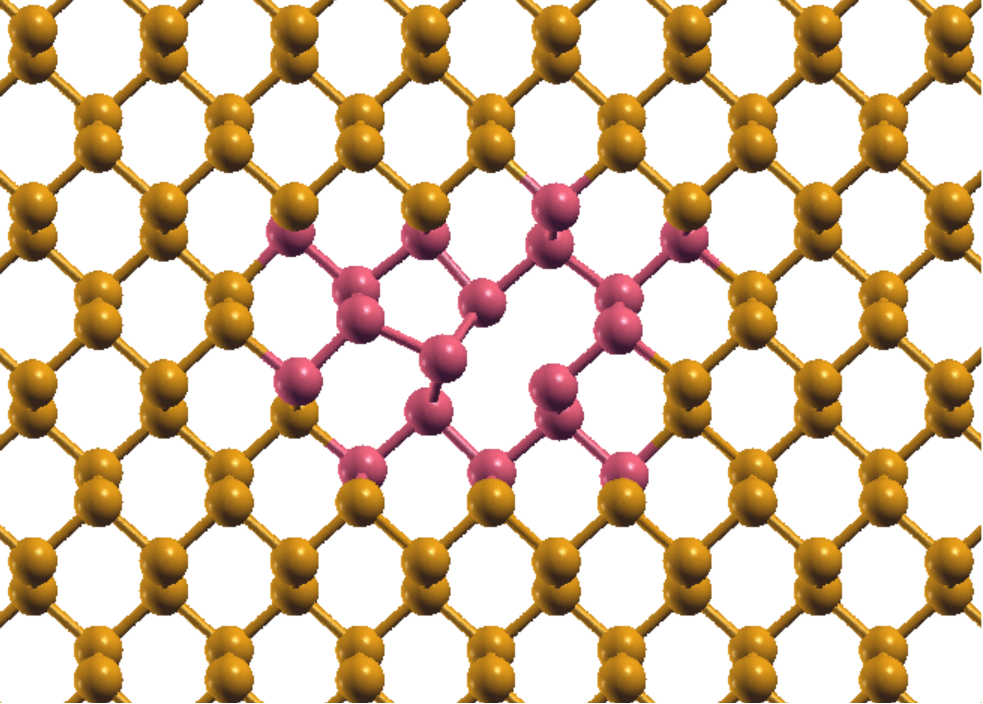}\\
\caption{The 18-atom fragment containing the negatively charged vacancy in phosphorene. Atoms belonging to the fragment used in the ADM are shown in pink, while atoms of the embedding environment are shown in gold. For details of the geometry optimization, see Ref. \onlinecite{Phosph_2026}.}\label{fig:frag}
\end{figure}

In this work, we calculate the correlation contribution to the formation energy as the energy of the defective fragment plus the energy of an isolated phosphorus atom in its quartet ground state, minus the energy of the corresponding fragment in the pristine structure; see Eq. (6) of Ref. \onlinecite{Phosph_2026}. We employ an 18-atom fragment from the defective structure, shown in Fig. \ref{fig:frag}, which corresponds to a 19-atom fragment for the pristine structure. The structural and computational parameters for the ADM calculations and the generation of the FCIDUMP files are the same as in Ref. \onlinecite{Phosph_2026}, including the POB-TZVP-rev2 basis set.\cite{VilelaOliveira2019} The post-HF methods, up to SVD-DC-CCSDT, are run with \elemco{} using the FCIDUMP-based input shown in Section \ref{sec:usage}. With the frozen-core approximation employed, the 18-atom fragment of the defective structure contains 76 electrons in 306 orbitals, while the 19-atom fragment of the pristine structure contains 80 electrons in 323 orbitals. The electron counts are smaller than the corresponding numbers of atomic valence electrons because some of the bonding electrons of the fragment-edge atoms belong to localized orbitals assigned to the environment and are therefore excluded from the fragment.\cite{Phosph_2026}
%Detailed orbital-space sizes, as well as timings for the different steps of the calculations, are given in the Supplementary Material.

For the phosphorus atom, we perform the complete calculation on the \elemco{} side using the same orbital and auxiliary basis sets as in the ADM calculations. The atom is surrounded by 11 ghost atoms taken from the pristine structure to account for basis-set superposition error. The quartet state is treated within the unrestricted formalism. For UDC-CCSDT, the SVD approximation is not applied because it is not implemented for the open-shell case.

\begin{table}
  %\centering
  \caption{Correlation-energy contribution $E^{\rm form}_{\rm corr}$ to the defect formation energy at different quantum-chemical levels and the corresponding timings $t$ (total wall time) on a 44-CPU node. For the SVD-based calculations, decomposition thresholds $\varepsilon_{T3}=10^{-5}$ and $10^{-6}$ were employed. The SVD error $\delta^{\rm SVD-(T)}$ is evaluated at the (T) level as $\delta^{\rm SVD-(T)}=E^{\rm form}_{\rm (T)}-E^{\rm form}_{\rm SVD-(T)}$. The post-CCSD(T) formation-energy contribution $\Delta E^{\rm post-CCSD(T)}$ is defined as $\Delta E^{\rm post-CCSD(T)}=E^{\rm form}_{\rm SVD-DC-CCSDT}-E^{\rm form}_{\rm SVD-CCSD(T)}$. The timings correspond to the calculations on the 19-atom fragment for the pristine structure (80 electrons in 323 orbitals) on a 12-core Intel Xeon Gold
6128 node with a rate of 3.40 GHz.}\label{tab:E_form}
  \begin{ruledtabular}
    \begin{tabular}{lcccc}
    Method & \multicolumn{2}{c}{$E^{\rm form}_{\rm corr}$, eV}&\multicolumn{2}{c}{$t$, h}\\
    \hline
%    \multicolumn{3}{c}{ Conventional treatment}\\
   MP2           &\multicolumn{2}{c}{1.061 }&\multicolumn{2}{c}{0.0} \\
   CCSD          &\multicolumn{2}{c}{0.865 }&\multicolumn{2}{c}{0.5} \\
    CCSD(T)       &\multicolumn{2}{c}{0.930 }&\multicolumn{2}{c}{4.1} \\
    \hline
    &\multicolumn{4}{c}{ SVD treatment}   \\
    &  \multicolumn{2}{c}{$\varepsilon_{T3}=10^{-5}$} &  \multicolumn{2}{c}{$\varepsilon_{T3}=10^{-6}$} \\
    &$E^{\rm form}_{\rm corr}$, eV&$t$, h&$E^{\rm form}_{\rm corr}$, eV&$t$, h\\
    \hline
   SVD-CCSD(T)             &   0.915  &1.7&  0.929&4.3 \\
   $\delta^{\rm SVD-(T)}$      &   0.015 & &  0.001& \\
   SVD-DC-CCSDT            &   0.904  & 6.7&  0.918&88.9 \\
   $\Delta E^{\rm post-CCSD(T)}$&  {\bf -0.011} & & {\bf-0.011}& \\
    \end{tabular}
\end{ruledtabular}    
\end{table}

The calculated formation energies, or more precisely their correlation-energy contributions, are listed in Table \ref{tab:E_form}. First, we note that the post-CCSD(T) correction is rather small, both in absolute terms ($-0.011$ eV or $0.25$ kcal/mol) and relative to the magnitude of the formation energy itself. Furthermore, it is smaller than the estimated remaining uncertainty in the benchmark of Ref. \onlinecite{Phosph_2026} due to finite-size effects and basis-set incompleteness. This suggests that the CCSD(T) level is already sufficient for the required accuracy of $E^{\rm form}_{\rm corr}$ for the phosphorene vacancy.

As this is an illustrative application, we have not attempted to converge $\Delta E^{\rm post-CCSD(T)}$ with respect to fragment size and/or basis-set size. Nevertheless, its small magnitude already for the 18/19-atom fragments and the POB-VTZ-rev2 basis suggests that the converged value is unlikely to become significant.

As for the SVD approximation, with a decomposition threshold of $10^{-6}$, the SVD error in $E^{\rm form}_{\rm corr}$ becomes essentially negligible. With $\varepsilon_{T3}=10^{-5}$, the error in $E^{\rm form}_{\rm corr}$ is still appreciably small, although not as small as for $10^{-6}$. Importantly, the post-CCSD(T) correction looks fully converged already at $\varepsilon_{T3}=10^{-5}$. Remarkably, the computational time of SVD-DC-CCSDT with $\varepsilon_{T3}=10^{-5}$ is comparable to that of CCSD(T).
%The Supplementary Material contains more detailed timings for the individual computational steps, as well as the sizes of the various orbital spaces involved in these calculations.

%==============================================================================
\section{Conclusions and outlook}\label{sec:conclusions}
%==============================================================================

\elemco{} is a modern quantum chemistry package written in Julia for molecular 
electron-correlation calculations. It combines an intuitive, macro-driven user 
interface that facilitates both routine calculations and rapid method 
prototyping with a transparent and extensible code base. State-of-the-art 
infrastructure for solvers, tensor-decomposition techniques like DF
or SVD, wavefunction-data 
management, and interoperability with external software makes it a versatile 
platform for electronic-structure method development. At the same time, Julia 
enables high computational performance without sacrificing readability or ease
of development. For input generation and visualization of results, the 
companion program \jlmol{} is available.

As a result, \elemco{} provides a flexible platform for both 
electronic-structure method developers and practitioners applying 
quantum-chemical methods. \elemco{} is free and open-source software; 
its source code, documentation, and examples are available 
online at \url{https://github.com/fkfest/ElemCo.jl} and \url{https://elem.co.il} 
The package offers a broad range of methods, 
spanning HF and MP2 to CC, DC, CI, and EOM approaches in closed-shell, 
restricted-open-shell, and unrestricted formulations.

Importantly, several methods are made openly available through \elemco{} 
for the first time or are, to the best of our knowledge, not freely 
accessible in any other package. These include tensor-decomposed DC methods 
such as SVD-DC-CCSDT, two-determinant and 
fixed-reference DC models, and the CIPHI variant of the selected-CI method. Furthermore, 
support for non-Hermitian Hamiltonians across many methods makes \elemco{} 
a natural platform for transcorrelated electronic-structure theory.

Additionally, we demonstrated the superior accuracy of EOM-DCSD 
compared to EOM-CCSD. Both methods are available in closed-shell, 
restricted open-shell, and spin-unrestricted variants.

Development of \elemco{} is ongoing and encompasses new method development, 
performance optimization, expansion of the package's capabilities, and 
additional interfaces to external software. Current efforts include the 
completion of SVD-DCSD and related tensor-decomposed methods,
quasi-variational approaches, extension of various methods to include positron 
affinity calculations, improving MCSCF and CASSCF, parallelization, and the 
implementation of additional molecular properties and analytical gradients, and continued improvements 
in computational performance.

%-----------------------------------------------------------------------------------------------------------------------------------------------
% new section
%-----------------------------------------------------------------------------------------------------------------------------------------------

\section*{Supplementary Material}
The Supplementary Material contains the EOM-DCSD benchmark data discussed 
in Section~\ref{subsec:excited_states}, the geometry of phenol used in 
the performance calculations presented in Section~\ref{sec:performance}, as well as the raw data and orbital-space sizes for the vacancy formation-energy calculations in Section~\ref{sec:phosph}.
~

%==============================================================================
\begin{acknowledgments}
Financial support from the Max Planck Society is gratefully acknowledged.
C.R. gratefully acknowledges Studienstiftung des deutschen
Volkes for a Master’s scholarship and Tiemann Stiftung for financial 
support of research visits to the Max Planck Institute for Solid
State Research.
 \end{acknowledgments}

\section*{Author Declarations}
\subsection*{Conflict of Interest}
The authors have no conflicts to disclose.

\section*{Data Availability}
The \elemco{} source code and documentation are openly
available at \url{https://github.com/fkfest/ElemCo.jl} and \url{https://elem.co.il}.

%==============================================================================
\bibliography{references}
%==============================================================================

\end{document}